\documentclass[manuscript]{aastex}

\newcommand{\myemail}{tanaka@vega.ess.sci.osaka-u.ac.jp}

\shorttitle{Spectral evolution model of PWNe}
\shortauthors{S. J. Tanaka and F. Takahara}

\begin{document}

\title{A MODEL OF THE SPECTRAL EVOLUTION OF PULSAR WIND NEBULAE}

\author{Shuta J. Tanaka\altaffilmark{1} and Fumio Takahara}
\affil{Department of Earth and Space Science, Graduate School of Science, Osaka University, 1-1 Machikaneyama-cho, Toyonaka, Osaka 560-0043, Japan }

\altaffiltext{1}{e-mail: \myemail}

\begin{abstract}
We study the spectral evolution of PWNe taking into account the energy injected when they are young.
We model the evolution of the magnetic field inside a uniformly expanding PWN.
Considering time dependent injection from the pulsar and coolings by radiative and adiabatic losses, we solve the evolution of the particle distribution function.
The model is calibrated by fitting the calculated spectrum to the observations of the Crab Nebula at an age of a thousand years.
The spectral evolution of the Crab Nebula in our model shows that the flux ratio of TeV $\gamma$-rays to X-rays increases with time, which implies that old PWNe are faint in X-rays, but not in TeV $\gamma$-rays.
The increase of this ratio is because the magnetic field decreases with time and is not because the X-ray emitting particles are cooled more rapidly than the TeV $\gamma$-ray emitting particles.
Our spectral evolution model matches the observed rate of the radio flux decrease of the Crab Nebula.
This result implies that our magnetic field evolution model is close to the reality.
Finally, from the viewpoint of the spectral evolution, only a small fraction of the injected energy from the Crab Pulsar needs to go to the magnetic field, which is consistent with previous studies.
\end{abstract}


\section{INTRODUCTION}\label{intro}

A pulsar releases its rotational energy as a relativistic magnetized outflow called a pulsar wind.
The pulsar wind collides with the surrounding supernova ejecta, forms the termination shock, and creates a PWN \citep{kc84a}.
The acceleration of the pulsar wind particles occurs at the termination shock and the PWN consists of the magnetic field and the ultrarelativistic particles \citep{rg74, kc84b}.
Created PWN emits photons ranging from radio to TeV $\gamma$-rays via the synchrotron radiation and the inverse Compton scattering.
Current status of the theoretical models as well as the observational confrontations is reviewed by \citet{gs06}.

The Crab Nebula is one of the best studied PWN at almost all observable wavelengths including its central pulsar, called the Crab Pulsar.
Many studies have been made to explain the observed properties of the Crab Nebula as a typical PWN.
\citet{kc84a} studied the spatial structure of the Crab Nebula, assuming that it is a steady state object (KC model).
They found that the magnetization parameter $\sigma$, the ratio of the electromagnetic energy flux to the particle energy flux just upstream the termination shock, must be as small as 0.003 to explain the observed dynamical properties of the Crab Nebula.
\citet{aa96} succeeded to reconstruct the current observed broadband spectrum of the Crab Nebula by the use of the KC model.

The KC model is not fit to consider the evolution because it is a steady state model.
However, it is important to consider the spectral evolution of the Crab Nebula.
To explain the flux decrease rates of the Crab Nebula in radio and optical wavelengths \citep[e.g.][]{ar85, v07, s03}, we need to consider the spectral evolution.
Moreover, we need to understand the spectral evolution of PWNe in general.
Recent observations have found many PWNe which have a variety of characteristics in terms of age, expansion velocity, morphology, radiation spectrum and others.
Some of these characteristics may be understood by the spectral evolution of PWNe.

For example, several old PWNe which are faint in X-rays have been detected in TeV $\gamma$-rays and \citet{dd08} discussed the possibility that some of the TeV $\gamma$-ray sources without an X-ray counterpart may be old PWNe.
When we study the broadband spectrum of old PWNe, we need to take into account the evolution of the injected energy from the pulsar, such as the magnetic field and the relativistic particles.
For old PWNe, most of the energy inside the PWNe is the energy which was injected when they were young because the spin-down power of the pulsar decreases fast with time.

Several spectral evolution models of PWNe have been studied \citep[e.g.][]{zet08, det08, get09}.
Although they included the evolution of the energy injection from the pulsar, several issues still remain to be further clarified.
These studies assume different evolution model of the particle distribution inside the PWNe.
\citet{zet08} considered the escape of the particles from the PWN which has a fixed volume.
Because a PWN is expanding while it is confined in the expanding supernova ejecta, it is more realistic to regard that particles rather suffer from an adiabatic loss than they escape from the nebula.
\citet{zet08} and \citet{det08} used the conversion efficiency of the spin-down power to the particle energy $\sim 10-50 \%$ to normalize the observed flux.
In their model, the magnetic field inside the PWN does not relate to the spin-down power and the remaining $50-90 \%$ of the spin-down power is not explicitly considered.
\citet{aa96} has a similar problem of the flux normalization.
\citet{get09} did not compare their calculated spectrum with the observations because they were interested mainly in the dynamical evolution of the PWN.

In this paper, we revisit a spectral evolution model of PWNe, paying attention to the issues mentioned above.
We do not take into account effects of the spatial structure of the PWN because it is somewhat costly and too detailed to discuss the whole radiation spectrum.
In our model, the PWN is simply treated as an expanding uniform sphere.
The energy inside the PWN is injected from the pulsar spin-down energy, which is divided between the magnetic field and the relativistic particles with a constant ratio.
Our simple model can describe the observed basic features.
Of course, many details are not treated in this simplified model, such as the filamentary structures and the spatial variations of photon indices.

We study the spectral evolution of the Crab Nebula as the first application of our model.
The Crab Nebula can be used as a calibrator of our model when it will be applied to other PWNe in future. 
In section \ref{model}, we describe our model of the PWN evolution.
In section \ref{crab}, we apply this model to the Crab Nebula.
In section \ref{discussion_conclusion}, discussions and conclusions are made.

\section{THE MODEL}\label{model}
For the calculation of the spectral evolution, we need to specify the evolution of the magnetic field and the particle distribution function.
Here, we describe the assumptions of our model.

\subsection{Model of the Energy Injection}\label{injection}
In this paper, we assume that the PWN is a uniform sphere expanding at a constant velocity $v_{\rm{PWN}}$.
The assumption of the constant velocity is the easiest way to take into account the expansion of the PWN, although the real behavior of the expansion must be more complex.
\citet{get09} investigated the dynamical evolution of the PWN surrounded by supernova ejecta.
They showed that the PWN expands at an almost constant velocity at first and then it turns to shrink and bounces in a late phase of the evolution after about 10kyr from birth.
We consider that the age of the PWN is younger than 10kyr in this paper and that the constant velocity is a good assumption in this range.
The radius of the PWN $R_{\rm{PWN}}(t)$ at a time $t$ is given by
\begin{equation}\label{eq1}
R_{\rm PWN}(t) = v_{\rm PWN} t.
\end{equation}

For the components inside the PWN, we assume that the PWN is composed of the magnetic field and the relativistic electron-positron plasma, both of which are injected from the pulsar inside the PWN.
The evolution of the spin-down power $L(t)$ is given by
\begin{equation}\label{eq2}
L(t) = L_{\rm 0} \left( 1+\frac{t}{\tau_{0}} \right)^{-\frac{n+1}{n-1}},
\end{equation}
where $L_0$ is the initial spin-down power and $\tau_0$ is the initial spin-down time. Both parameters are fixed, if the current pulsar period $P$, its time derivative $\dot{P}$, braking index $n$ and age of the pulsar $t_{\rm{age}}$ are known, assuming that the moment of inertia of the pulsar is $10^{45}\rm g \cdot \rm cm^2$.

We divide the energy injection from the pulsar into the magnetic field energy $\dot{E}_{\rm{B}}$ and the relativistic particle energy $\dot{E}_{\rm{e}}$ using the time independent parameter $\eta$ ($0 \leq \eta \leq 1$).
The fraction parameter $\eta$ is the ratio of the magnetic field energy injection to the spin-down power, i.e.,
\begin{eqnarray}\label{eq3}
L(t) & = & \dot{E}_{\rm{e}}(t) + \dot{E}_{\rm{B}}(t) \nonumber \\
& = & (1 - \eta) \cdot L(t) + \eta \cdot L(t).
\end{eqnarray}
The fraction parameter $\eta$ in our model is similar to the magnetization parameter $\sigma$ in the KC model, although they are not the same.
The magnetization parameter $\sigma$ is the ratio $\dot{E}_{\rm{B}} / \dot{E}_{\rm{e}}$ at the pulsar wind region immediately upstream the termination shock.
On the other hand, the fraction parameter $\eta$ pertains to the ratio $\dot{E}_{\rm{B}} / (\dot{E}_{\rm{B}} + \dot{E}_{\rm{e}})$ into the PWN region.

For the particle injection, we also need to determine the injection spectrum of the relativistic particles.
Following \citet{vd06}, we assume that the injection spectrum of the relativistic particles $Q_{\rm inj}(\gamma, t)$ obeys a broken power-law
\begin{equation}\label{eq4}
Q_{\rm inj}(\gamma, t) = \left\{
\begin{array}{ll}
Q_{\rm 0}(t) (\gamma /\gamma_{\rm {b}})^{-p_{\rm 1}} & \mbox{ for $\gamma_{\rm min} \leq \gamma \leq \gamma_{\rm b}$ ,} \\
Q_{\rm 0}(t) (\gamma / \gamma_{\rm {b}})^{-p_{\rm 2}} & \mbox{ for  $\gamma_{\rm b} \leq \gamma \leq \gamma_{\rm max}$ ,}
\end{array} \right.
\end{equation}
where $\gamma$ is the Lorentz factor of the relativistic electrons and positrons.
We introduce time independent parameters $\gamma_{\rm min}$, $\gamma_{\rm b}$, $\gamma_{\rm max}$, $p_{\rm 1}$ and $p_{\rm 2}$ which are the minimum, break and maximum Lorentz factors and the power-law indices at the low and high energy ranges of the injection spectra, respectively.
Below, we use terms the low/high energy particles to refer to the particles which have the energy lower/higher than $\gamma_{\rm b}$.
Note that most of PWNe have value $p_{\rm 1} <2$ and $p_{\rm 2} >2$, which means that the energy of the particle injection is dominated by the particles $\gamma = \gamma_{\rm b}$.

We require that the normalization $Q_{\rm{0}}(t)$ satisfies the following equation
\begin{eqnarray}\label{eq5}
(1-\eta)L(t) = \int_{\gamma_{\rm min}}^{\gamma_{\rm max}} Q_{\rm inj}(\gamma, t) \gamma m_{\rm e} c^{2} d\gamma, 
\end{eqnarray}
where $m_{\rm e}$ and $c$ are the mass of an electron (or positron) and the speed of light, respectively.
From equations (\ref{eq1}), (\ref{eq4}), and (\ref{eq5}), the normalization parameter $Q_{\rm 0}(t)$ is given by
\begin{equation}\label{eq6}
Q_{\rm 0}(t) = \frac{L_{\rm 0} \cdot (1-\eta)}{m_{\rm e}c^{2}} \cdot \left(1+\frac{t}{\tau_{\rm 0}} \right)^{-\frac{n+1}{n-1}} 
\left(\frac{\gamma_{\rm b}^{2}(p_{\rm 1}-p_{\rm 2})}{(2-p_{\rm 1})(2-p_{\rm 2})} + \frac{\gamma_{\rm b}^{p_{\rm 2}} \cdot \gamma_{\rm max}^{2-p_{\rm 2}}}{2-p_{\rm 2}} - \frac{\gamma_{\rm b}^{p_{\rm 1}} \cdot \gamma_{\rm min}^{2-p_{\rm 1}}}{2-p_{\rm 1}} \right)^{-1} .
\end{equation}

\subsection{Model of the Evolution of the Magnetic Field}\label{mag_evol}
Because the magnetic field lines are stretching and winding, it is difficult to model the evolution of the magnetic field $B(t)$ in the context of the uniform PWN.
We have to solve the relativistic magnetohydrodynamics (MHD) equations to determine the realistic magnetic field evolution \citep[e.g.][]{det04}.
In this paper, for simplicity, we assume that the magnetic field evolution is determined in the form of the magnetic field energy conservation, i.e.,
\begin{eqnarray}\label{eq7}
\frac{4\pi}{3} (R_{\rm{PWN}}(t))^{3} \cdot \frac{(B(t))^{2}}{8\pi} & = & \int_0^{t} \eta L(t') dt' \nonumber \\
& = & \eta E_{\rm spin}(t),
\end{eqnarray}
where $E_{\rm spin}(t)$ is the integrated spin-down energy at a time $t$.
From equations (\ref{eq1}), (\ref{eq2}) and (\ref{eq7}), the magnetic field at a time $t$ is given by
\begin{equation}\label{eq8}
B(t) = \sqrt{\frac{3 (n-1) \eta L_0 \tau_0}{(R_{\rm PWN})^3} \left\{ 1 - \left( 1+ \frac{t}{\tau_{\rm 0}} \right)^{-\frac{2}{n-1}} \right\} }.
\end{equation}
The magnetic field approximately evolves as $B(t) \propto t^{-1.5}$ for $t > \tau_0$.
Note that although this magnetic field evolution model may be ad hoc, its behavior is very similar to those adopted in other works as we compare our model with other representative treatments \citep{rg74, kc84a, det09, vd06}.

\citet{rg74} considered that the magnetic field evolution is determined by the number of turns of the central pulsar.
They considered that the magnetic field in the PWN was built up by the winding of field lines because of the pulsar spin.
The total number of the turns of the magnetic field line is given by
\begin{equation}\label{eq9}
N(t) = \int^t_{t_{\min}} \frac{\Omega(t')}{2 \pi} dt'.
\end{equation}
This number is proportional to the magnetic flux, which means
\begin{equation}\label{eq10}
B(t) \propto N(t) (R_{\rm PWN}(t))^{-2}.
\end{equation}
For $t > \tau_0$, we obtain $B(t) \propto t^{n / (1-n)}$ so that $B(t) \propto t^{-1.5}$ for $n = 3$ and $B(t) \propto t^{-5 / 3}$ for $n = 2.5$.
Our choice of the magnetic field evolution is not so different.

\citet{kc84a} considered the steady state spatial structure of PWN.
In the KC model, the magnetic field increases with the distance from the pulsar till the magnetic pressure dominates over the particle pressure.
Although the assumption of the steady state PWN makes it difficult to compare with our model, it seems to be natural to regard that the magnetic energy is a constant fraction of the total energy ($\sim$ the particle energy), which means
\begin{equation}\label{eq11}
(B(t))^2 (R_{\rm PWN}(t))^3 \propto \frac{E_{\rm spin}(t)}{R_{\rm PWN}(t)},
\end{equation}
where we include the adiabatic cooling of the total energy, so that $B(t) \propto t^{-2}$ for $t > \tau_0$.
Equation (\ref{eq11}) is alternatively interpreted as the magnetic flux conservation.
The magnetic field decreases more rapidly than our model, but it is still close to our model.

In \citet{det09}, they mentioned that the spatially averaged magnetic field strength of the PWN decreases as $B(t) \propto t^{-1.3}$ in their calculation of non-relativistic MHD equations.
This is again close to our model.

Finally, \citet{vd06} assumed that the magnetic field of the PWN evolved as $B(t) = B_0 / (1 + (t / T)^k)$ and \citet{zet08} used this model with $k = 0.5$, i.e., $B(t) \propto t^{-0.5}$ for $t > T = 500 \rm{yr}$.
This is somewhat different from other models presented above and from ours.
Based on these comparisons, we regard that our model is close to the reality, although other choices are also possible.

\subsection{Model of the Evolution of the Particle Distribution}\label{part_evol}
We assume that the distribution of the particles in the PWN is isotropic, so that the particle distribution function can be easily volume integrated to be described by the energy distribution $N(\gamma, t)$.
The evolution of the particle distribution $N(\gamma, t)$ is given by the continuity equation in the energy space,
\begin{equation}\label{eq12}
\frac{ \partial}{ \partial t} N(\gamma, t) + \frac{ \partial}{ \partial \gamma} \left( \dot{\gamma}(\gamma, t) N(\gamma, t) \right) = Q_{\mathrm{inj}}(\gamma, t). 
\end{equation}

We consider the cooling effects of the relativistic particles $\dot{\gamma}(\gamma, t)$ including the synchrotron radiation, the inverse Compton scattering off the Cosmic Microwave Background Radiation (CMB) and the adiabatic expansion, i.e.,
\begin{equation}\label{eq13}
\dot{\gamma}(\gamma, t) = \dot{\gamma}_{\mathrm{syn}}(\gamma,t) + \dot{\gamma}_{\mathrm{IC}}(\gamma) + \dot{\gamma}_{\mathrm{ad}}(\gamma,t).
\end{equation}
The inverse Compton cooling $\dot{\gamma}_{\mathrm{IC}}(\gamma)$ does not depend on time because we consider that the target photon field is only the CMB.
As the cooling effect, we do not include the inverse Compton scattering off the synchrotron radiation field because it never be a more important cooling process than the synchrotron cooling and because it costs too much computer power.
Note that we include it in the calculation of the radiation spectrum, as described in section \ref{spectrum}. 

The synchrotron cooling $\dot{\gamma}_{\mathrm{syn}}(\gamma,t)$ is given by \citep[e.g.][]{rl79}
\begin{equation}\label{eq14}
\dot{\gamma}_{\mathrm{syn}}(\gamma,t) = - \frac{4}{3} \cdot \frac{\sigma_{\rm{T}}^{}}{m_{\rm{e}} c} \cdot U_{\mathrm{B}}(t) \gamma^2 ,
\end{equation}
where $\sigma_{\mathrm{T}}$ is the Thomson cross section and $U_{\mathrm{B}}(t)$ is the magnetic field energy density calculated from equation (\ref{eq8}).

The inverse Compton cooling $\dot{\gamma}_{\mathrm{IC}}(\gamma)$ is given by \citep[e.g.][]{bg70} 

\begin{equation}\label{eq15}
\dot{\gamma}_{\mathrm{IC}}(\gamma) = - \frac{3}{4} \cdot \frac{\sigma_{\mathrm{T}}^{} h}{m_{\mathrm{e}} c} \cdot \frac{1}{\gamma^2} \int_0^{\infty} \nu_{\mathrm{fin}} d\nu_{\mathrm{fin}} \int_0^{\infty} \frac{n_{\mathrm{CMB}}(\nu_{\mathrm{ini}})}{\nu_{\mathrm{ini}}} \cdot f(q, \Gamma_{\rm{\epsilon}}) \cdot \theta(1-q) \cdot \theta(q-1/4\gamma^2) d\nu_{\mathrm{ini}} ,
\end{equation}
where $h$ is the Planck's constant, $\nu_{\rm{ini}}$ and $\nu_{\rm{fin}}$ are the frequency of the CMB photons and that of scattered photons, respectively, $n_{\rm{CMB}}(\nu_{\rm{ini}})$ is the distribution of the CMB described in equation (\ref{eq19}) below.
$\Gamma_{\rm{\epsilon}} = 4 \gamma h \nu_{\rm{ini}} / (m_{\rm{e}} c^2)$, $q = h \nu_{\rm{fin}} /(\Gamma_{\rm{\epsilon}} (\gamma m_{\rm{e}} c^2 - h \nu_{\rm{fin}}))$, $f(q, \Gamma_{\rm{\epsilon}}) = 2q \ln q + (1+2q)(1-q) + 0.5 (1-q) (\Gamma_{\epsilon}q)^2 / (1+\Gamma_{\epsilon}q)$, and $\theta$ is the step function.

Finally, the adiabatic cooling $\dot{\gamma}_{\mathrm{ad}}(\gamma, t)$ is given by
\begin{eqnarray}\label{eq16}
\dot{\gamma}_{\mathrm{ad}}(\gamma, t) & = & - \frac{v_{\mathrm{PWN}}}{R_{\rm PWN}(t)} \cdot \gamma \nonumber \\
& = & - \frac{\gamma}{t},
\end{eqnarray}
where we use equation (\ref{eq1}).
Note that the adiabatic cooling is independent of the expansion velocity $v_{\rm PWN}$.

As was stated in section \ref{intro}, we consider that it is more reasonable to treat the adiabatic loss rather than the escape loss.
\citet{zet08} considered an escape of the particles instead of an adiabatic loss.
They treated the escape of the particles based on the Bohm diffusion.
Because higher energy particles have larger Lamor radii, the high energy particles suffer from the escape loss in their models.
On the other hand, the adiabatic loss is the dominant cooling process for the low energy particles in our model, since the high energy particles suffer from stronger radiative loss.

\subsection{Calculation of Spectrum of the PWN}\label{spectrum}
When the evolutions of the magnetic field and the particle distribution are determined, we can calculate the spectral evolution of the PWN.
To compare the calculated luminosity with the observed flux, we assume that the radiation is isotropic.
The radiation processes which we consider in our model are the synchrotron radiation and the inverse Compton scattering off the CMB (IC/CMB) and the synchrotron radiation (SSC).

The synchrotron radiation luminosity $L_{\nu,\rm syn}(t)$ is given by \citep[e.g.][]{rl79}
\begin{equation}\label{eq17}
L_{\nu,\rm syn}(t) = \int_{\gamma_{\rm min}}^{\gamma_{\rm max}} N(\gamma, t) P(\nu, \gamma, B(t)) d\gamma ,
\end{equation}
where $P(\nu, \gamma, B(t)) = \sqrt{3} e^3 B(t) / (m_{\rm e} c^2) F(\nu / \nu_{\rm c}(t))$ is the emissivity of the synchrotron radiation per particle, $\nu_{\rm c}(t) = 3 e B(t) \gamma^2 / (4 \pi m_{\rm e} c)$, $F(x) = x \int^{\infty}_x K_{5/3}(y) dy$ with $x = \nu / \nu_{\rm{c}}$ and $K_{5/3}$ being the modified Bessel function of order $5/3$.

The inverse Compton scattering luminosity $L_{\nu_{\rm fin},\rm IC}(t)$ is given by \citep[e.g.][]{bg70}
\begin{equation}\label{eq18}
L_{\nu_{\rm fin},\rm IC}(t) = \frac{3}{4} \cdot \frac{\sigma_{\rm T}^{} h \nu_{\rm fin} }{m_{\rm e} c} \int_{\gamma_{\rm min}}^{\gamma_{\rm max}} \frac{N(\gamma, t)}{\gamma} d\gamma \int_0^{\infty} \frac{n_{\rm ph}(\nu_{\rm ini})}{\nu_{\rm ini}} \cdot f(q, \Gamma_{\rm \epsilon}) d\nu_{\rm ini} ,
\end{equation}
where $n_{\rm{ph}}(\nu_{\rm ini})$ is the distribution of the target photon fields including the CMB and the synchrotron radiation.
The distribution of the CMB $n_{\rm CMB}^{}(\nu)$ is given by the black-body distribution
\begin{equation}\label{eq19}
n_{\rm CMB}(\nu) = \frac{8\pi}{c^3} \cdot \frac{\nu^2}{\exp{(h \nu/ k_{\rm{B}} T_{\rm{CMB}})}-1} ,
\end{equation}
where $k_{\rm B}$ is the Boltzmann's constant and $T_{\rm CMB}$ is the temperature of the CMB.
The distribution of the synchrotron radiation $n_{\rm syn}(\nu, t)$ is given by
\begin{equation}\label{eq20}
n_{\rm syn}(\nu, t) = \frac{L_{\nu,\rm syn}(t)}{4\pi R^2_{\rm PWN}(t) c} \cdot \frac{1}{h \nu} \cdot \overline{U} ,
\end{equation}
where $\overline{U} \sim 2.24$ is the average of $U(x) = (3 / 2) \int^1_0 (y / x) \ln{((x+y)/(x-y))} dy$ in a spherical volume and $x$ is the ratio of the distance from the center of the PWN to the PWN radius $R_{\rm PWN}(t)$ \citep[c.f.][]{aa96}.

In equation (\ref{eq20}), \citet{zet08} did not use $R_{\rm PWN}(t)$, but a new parameter $r_{\rm syn}$ ($< R_{\rm PWN}(t)$), considering that the spread of the synchrotron radiation field is smaller than the radius of the PWN.
Because only the synchrotron photons whose frequency below optical wavelengths contribute to the SSC flux and because the observed size of the optical synchrotron nebula is comparable to the size of the PWN, we use $R_{\rm PWN}(t)$ in equation (\ref{eq20}).

\section{APPLICATION TO THE CRAB NEBULA}\label{crab}
In this section, we apply our model to the Crab Nebula as the standard calibrator of our model.
The Crab Nebula is one of the best observed PWN in almost all observable wavelengths.
The results of the application to the Crab Nebula will be a landmark for applications of our model to other PWNe.

First, we determine the parameters to reproduce the current Crab Nebula.
We can obtain the information about the magnetic field and the particle distribution and compare with those of the steady state solution of \citet{aa96}.
Secondly, we show the spectral evolution of the Crab Nebula in our model with the use of the determined parameters.
Through the evolutions of the magnetic field and the particle distribution, we explain how the behavior of the spectral evolution can be understood.
We also show that our spectral evolution model is in a reasonably good agreement with the observations of the radio/optical flux decreases.
Thirdly, we give a simple argument to see the dependences on the fraction parameter $\eta$ of the spectral evolution.
Based on the general properties of PWNe in this simple argument, we can apply this to other PWNe in future works.
Finally, we discuss about the fitted parameters other than the fraction parameter $\eta$, which characterize the injection spectrum (equation (\ref{eq4})).

The Crab pulsar has the period $3.31 \times 10^{-2}\rm s$, its time derivative $4.21 \times 10^{-13} \rm {s} \cdot \rm {s}^{-1}$ and braking index 2.51.
The progenitor supernova is SN1054, which means the age of the Crab Nebula $t_{\rm age} \sim 950\rm yr$.
Because all the necessary pulsar parameters are known, the evolution of the spin-down power in equation (\ref{eq2}) is fixed, as $\tau_0 \sim 700 \rm yr$ and $L_0 = 3.4 \times 10^{39} \rm ergs \cdot s^{-1}$.
The distance to the Crab Nebula $2\rm kpc$ is used to convert the luminosity into the isotropic flux.
The Crab synchrotron Nebula is roughly an elliptical shape with a major axis of $4.4\rm pc$ and a minor axis of $2.9\rm pc$.
Here, we regard that the Crab Nebula is a sphere of the diameter $\sim 3.5\rm pc$.
Combining with $t_{\rm age} \sim 950\rm yr$, the constant expansion velocity becomes $v_{\rm PWN} \sim 1800 \rm km/s$, which is close to the observed expansion velocity of the Crab Nebula.

\subsection{Current Spectrum of the Crab Nebula}\label{crab_current}
Figure \ref{f1} shows calculated current spectrum of the Crab Nebula with the current observational data.
The adopted parameters are shown in Table \ref{tbl-1}.
As seen in Figure \ref{f1}, the SSC flux is stronger than the IC/CMB flux in $\gamma$-rays.
Note that the IC/CMB flux is almost in the Thomson regime, while the SSC flux is largely affected by the Klein-Nishina effect.

We fit the data with the parameters $\eta = 0.005$, $\gamma_{\rm max} = 7.0 \times 10^9$, $\gamma_{\rm b} = 6.0 \times 10^5$, $\gamma_{\rm min} = 1.0 \times 10^2$, $p_1 = 1.5$, and $p_2 = 2.5$.
The fraction parameter $\eta$ governs the absolute values of the fluxes and the flux ratio of the inverse Compton scattering to the synchrotron radiation, as discussed in more detail in section \ref{explanation}.
The KC model derived $\sigma \ll 1$ from the viewpoint of the current dynamical structure of the Crab Nebula, while we determine $\eta \ll 1$ from the viewpoint of the spectral evolution.
The parameters $\gamma_{\rm max}$, $\gamma_{\rm b}$, $p_1$, and $p_2$ are fixed to reproduce the observed synchrotron spectral shape, such as the spectral breaks and the photon indices, while $\gamma_{\rm min}$ should be regarded as an upper limit to reproduce the radio flux at the lowest frequency.
In section \ref{other}, these fitted parameters characterizing particle injection are discussed in detail. 

In our calculation, the current magnetic field strength of the Crab Nebula turns out to be $B_{\rm now} = 85\mu \rm{G}$, which is smaller than $\sim 300 \mu \rm{G}$ used by \citet{aa96}.
This difference of the magnetic field strength can be explained as follows.
\citet{aa96} adopted $B_{\rm KC} \sim 300 \mu \rm G$ from the KC model and adjusted the particle number to reproduce the observations.
They applied roughly half a spin-down power compared with the KC model to reproduce the spectrum and thus the other half is missing.
On the other hand, all the injected spin-down power is divided between the magnetic field and the particle energies in our model.
If we adopt $B_{\rm now} = B_{\rm KC} \sim 300 \mu \rm G$, the synchrotron flux and also the SSC flux increase by about an order of magnitude.
Note that the relativistic MHD simulation by \citet{vet08} also indicates a smaller value of the spatially averaged magnetic field strength $ \sim 100 \mu G$, which is close to our value $B_{\rm now} = 85\mu \rm{G}$.

\subsection{Spectral Evolution of the Crab Nebula}\label{crab_evolution}
Our model can calculate the past and the future spectra of the Crab Nebula.
All the calculated results in this section shown in Figures \ref{f2}-\ref{f5} are with the use of the same parameters as Figure \ref{f1}.

Figure \ref{f2} shows the spectral evolution.
As seen in Figure \ref{f2}, the synchrotron flux decreases with time and the SSC flux also decreases with time in accordance with the synchrotron flux, while the IC/CMB flux decreases more slowly than the SSC flux.
We discuss their time dependence in section \ref{explanation}.
An important result in Figure \ref{f2} is that the flux ratio of $\gamma$-rays to X-rays increases with time.
This supports the view that old PWNe can be observed as $\gamma$-ray sources with no or a weak X-ray counterpart.

The radio/optical observations of the Crab Nebula have suggested that the radio/optical flux of the Crab Nebula is decreasing with time.
The inferred rate of the radio flux decrease is $- 0.17 \pm 0.02 \% / \rm yr$ and almost independent of the frequency for a range 86 - 8000MHz \citep{v07}.
Our model predicts the current rate $\sim - 0.16\% / \rm yr$, which is almost consistent with the observation.
The inferred rate of the optical continuum flux decrease is $- 0.55 \% / \rm yr$ calibrated at 5000\AA \ \citep{s03}.
Our model predicts the current rate $\sim - 0.24 \% / \rm yr$, which is a factor of two smaller than the observation.
However the trend that the decreasing rate increases with frequency matches the observations.
This is because the optical emitting particles suffer from stronger synchrotron cooling than the radio emitting particles. 

To understand the detailed features shown in Figure \ref{f2}, we examine features of the evolution of the particle energy distribution in Figure \ref{f3}.
For comparison, the injected spectrum of the particles till 10kyr without the cooling effects is plotted by the dot-dashed line in Figure \ref{f3}. 
The evolution of the particle distribution is characterized differently in four energy ranges.
First, for $\gamma > 10^8$, the particle number increases with time.
This increase of the high energy particles relates to an increase of the IC/CMB flux in Figure \ref{f2} after 1kyr above $10 \rm TeV$. 
Secondly, for $10^5 < \gamma < 10^8$, the behavior of the particle distribution is complex.
The particle number and the power-law index in this energy range do not monotonically change.
The radiation spectrum in the range from infrared to X-rays reflects this complex behavior.
Thirdly, for $10^2 < \gamma < 10^5$, the change of the particle distribution is small.
The difference between all the lines for several epochs is less than a factor of two.
This leads to an important conclusion that the radio flux decrease is mainly due to the decrease of the magnetic field, so that the observations of the radio flux decrease support our model of the magnetic field evolution.
Lastly, for $\gamma < 10^2$, there exist particles with $\gamma < \gamma_{\rm{min}}$.
This is because the adiabatic cooling is still effective at low energy.
We discuss more about the features of the particle $\gamma > 10^2$ in later paragraphs, which relate to the spectral evolution above $10^7 \rm Hz$.
As discussed below, these features indicate the importance to consider the evolution of the particle injection rate and the cooling time due to various processes.

Figure \ref{f4} plots the evolution of cooling time $\tau(\gamma, t) = \gamma / \vert \dot{\gamma}(\gamma, t) \vert$ and helps to understand features of the particles with $\gamma > 10^2$ in Figure \ref{f3}. 
As seen in Figure \ref{f4}, at an age of 1kyr, the synchrotron cooling is effective for the particles with $\gamma > 10^6$, while the adiabatic cooling is effective with $\gamma < 10^6$.
However, because of a rapid decrease of the magnetic field energy density, the adiabatic cooling dominates at the $\gamma < 10^8$ at an age of 10kyr.
From equations (\ref{eq8}), (\ref{eq14}) and (\ref{eq16}), $\tau_{\rm syn} \propto \gamma^{-1} t^{2}$ for $t < \tau_{0}$ and $\tau_{\rm syn} \propto \gamma^{-1} t^{3}$ for $t > \tau_{0}$, while $\tau_{\rm ad} \propto t$ at any time.

We examine the increase of the particle number for $\gamma > 10^8$ shown in Figure \ref{f3}.
The increase means that the particle injection dominates the cooling effect in this energy range.
Equation (\ref{eq12}) is approxmately expressed as ($\partial / \partial t \sim t^{-1} = \tau^{-1}_{\rm dyn}$ and $\partial \dot{\gamma} / \partial \gamma \sim \tau^{-1}_{\rm cool} = (\tau^{-1}_{\rm syn} + \tau^{-1}_{\rm IC} + \tau^{-1}_{\rm ad})$)
\begin{equation}\label{eq21}
N(\gamma, t) \sim \frac{\tau_{\rm dyn} \tau_{\rm cool}}{\tau_{\rm dyn} + \tau_{\rm cool}} \cdot Q_{\rm{inj}}(\gamma, t).
\end{equation}
Because synchrotron cooling dominates in this energy range $\tau_{\rm cool} \sim \tau_{\rm syn} < \tau_{\rm dyn}$, equation (\ref{eq21}) becomes $N \sim \tau_{\rm syn} Q_{\rm{inj}}$.
For $t < \tau_0$, the injection term behaves as $Q_{\rm{inj}} \propto \gamma^{- p_2}$, so that $N \propto \gamma^{- p_2 - 1} t^2$.
For $t > \tau_0$, the injection term behaves as $Q_{\rm{inj}} \propto \gamma^{- p_2} t^{-7/3}$ for $n = 2.5$, so that $N \propto \gamma^{- p_2 - 1} t^{2/3}$.
This is the reason why the particle number increases with time and the particle distribution is softer than the injection distribution for $\gamma > 10^8$.

We next examine the steadiness of the particles number for $10^2 < \gamma < 10^5$ shown in Figure \ref{f3}.
For $t < \tau_0$, because $\tau_{\rm cool} \sim \tau_{\rm ad} \sim \tau_{\rm dyn}$ in equation (\ref{eq21}), we have $N \propto \gamma^{- p_1} t$.
For $t > \tau_0$, because the injection term more rapidly decreases with time than the adiabatic cooling term, equation (\ref{eq12}) is approximated by 
\begin{equation}\label{eq22}
\frac{ \partial}{ \partial t} N(\gamma, t) \sim \frac{ \partial}{ \partial \gamma} \left( \frac{\gamma}{t} N(\gamma, t) \right).
\end{equation}
When we assume a form $N \propto \gamma^{-\alpha} t^{\beta}$, we obtain $\beta = 1 - \alpha$.
For $\alpha = p_1 = 1.5$, we have $\beta = - 0.5$.
The particle number increases until $t \sim \tau_0$, and then it decreases slowly with time as $N \propto t^{-0.5}$ for $10^2 < \gamma < 10^5$. 

The feature of the particle distribution for $10^5 < \gamma < 10^8$ shown in Figure \ref{f3} is complex because the dominant cooling process changes with time as seen in Figure \ref{f4}.
Until an age of a few thousand years, the feature is similar to that for $\gamma > 10^8$ (equation (\ref{eq21}) with $\tau_{\rm cool} \sim \tau_{\rm syn}$), i.e., the particle number increases with time and the particle distribution is softer than the injection distribution.
After that, the feature becomes similar to that for $10^2 < \gamma < 10^5$ for $t > \tau_0$ (equation (\ref{eq22})) and the particle number decreases with time.
Because the time when $\tau_{\rm ad} \sim \tau_{\rm syn}$ is different for different energy, the particle distribution becomes harder than $\propto \gamma^{-p_2 - 1}$ at an age of 10kyr.

In Figure \ref{f5}, we show the evolution of each energy component: the particle energy, the magnetic field energy, the radiated energies via the synchrotron radiation and the inverse Compton scattering and the wasted energy via the adiabatic expansion, all of which are normalized by the integrated spin-down energy $E_{\rm spin}(t)$.
As seen in Figure \ref{f5}, although almost all the injected energy goes to the particle energy, the particle energy decreases with time by the cooling effects.
The magnetic field energy is always constant fraction of the injected energy $\eta E_{\rm spin}(t)$ as assumed in equation (\ref{eq7}).
Most of the particle energy is cooled by the synchrotron radiation in the early phase ($< 1\rm kyr$) because the cooling break due to the synchrotron cooling is a little smaller than $\gamma_{\rm b}$ as seen in Figure \ref{f3}.
On the other hand, most of the injected energy goes to the adiabatic loss in the late phase ($> 1\rm kyr$), which means that most of the injected energy goes to the kinetic energy of the supernova ejecta.
The inverse Compton cooling is not important in the case of the Crab Nebula for $t < 10\rm kyr$, which is also seen in Figure \ref{f4}.
Note that the integrated spin-down energy $E_{\rm spin}(t)$ also increases with time, for example, $E_{\rm spin}(100\rm yr) = 9.1 \times 10^{48} \rm ergs$, $E_{\rm spin}(1\rm kyr) = 3.9 \times 10^{49} \rm ergs$ and $E_{\rm spin}(10\rm kyr) = 5.5 \times 10^{49} \rm ergs$.

\subsection{Characteristics of the Spectral Evolution}\label{explanation}
We discuss the evolution of powers of the synchrotron radiation and the inverse Compton scattering and their dependence on the fraction parameter $\eta$, which will help to apply our model to other PWNe in future.

We assume $t > \tau_0$ in this section aiming to discuss primarily old PWNe.
For simplicity, we also assume that the particle energy is expressed as $N(\gamma_{\rm b}, t) \gamma^2_{\rm b} m_{\rm e} c^2 = (1-\eta) \xi(t) E_{\rm spin}(t)$, where $\xi(t) < 1$ accounts for the cooling effects.
As seen in Figure \ref{f5}, $N(\gamma_{\rm b}, t) \gamma^2_{\rm b} m_{\rm e} c^2 \sim \rm const.$ and $\xi(t) \sim \rm const. \sim 0.1$ is a fairly good approximation.
Moreover, we assume that powers of all radiation mechanisms are dominated by the emission from the particles $\gamma = \gamma_{\rm b}$.
This assumption is valid when the low energy power-law index $< 2$ and the high energy power-law index $> 3$ in the particle distribution.

The power ratio of the inverse Compton scattering to the synchrotron radiation is given by

\begin{equation}\label{eq23}
\frac{P_{\rm IC}(t)}{P_{\rm syn}(t)} \sim f_{\rm KN} \frac{U_{\rm ph}(t)}{U_{\rm B}(t)},
\end{equation}
where $f_{\rm KN} < 1$ represents the Klein-Nishina effect and $U_{\rm ph}(t)$ is the energy density of the target photon field.
$f_{\rm KN} \sim 1$ for the IC/CMB and $f_{\rm KN} < 0.1$ for the SSC.

The power of the synchrotron radiation is given by \citep[e.g.][]{rl79}

\begin{eqnarray}\label{eq24}
P_{\rm {syn}}(t) & \sim & \frac{4}{3} \sigma_{\rm {T}} c \gamma^2_{\rm {b}} U_{\rm {B}}(t) \cdot \gamma_{\rm {b}} N(\gamma_{\rm {b}}, t) \nonumber \\ 
& = & \frac{\sigma_{\rm {T}} \gamma_{\rm b}}{\pi m_{\rm {e}} c} \cdot \frac{\eta (1 - \eta)}{R^3_{\rm PWN}(t)} \xi(t) E_{\rm spin}^2(t) \propto \eta (1 - \eta) t^{-3},
\end{eqnarray}
where  $\gamma_{\rm b} N(\gamma_{\rm b}, t)$ is the number of the particles around $\gamma = \gamma_{\rm b}$ and we use equation (\ref{eq7}).
Note that although the power of the synchrotron radiation has been conventionally compared with the spin-down power $L(t)$, the power of the synchrotron radiation relates to the integrated spin-down energy $E_{\rm spin}(t)$ rather than the instantaneous spin-down power $L(t)$ in equation (\ref{eq24}).

Now equation (\ref{eq23}) becomes, in the case of the IC/CMB ($U_{\rm ph}(t) = U_{\rm CMB}$), 

\begin{eqnarray}\label{eq25}
\frac{P_{\rm IC/CMB}(t)}{P_{\rm syn}(t)} & \sim & f_{\rm KN} \left( \frac{B(t)}{3 \mu G} \right)^{-2}  U_{\rm CMB} \nonumber \\
& \propto & \frac{R^3_{\rm PWN}(t)}{\eta E_{\rm spin}(t)} \propto \eta^{-1} t^3.
\end{eqnarray}
In the case of the SSC, we estimate the synchrotron radiation energy density $U_{\rm syn}(t)$ as

\begin{equation}\label{eq26}
U_{\rm syn}(t) \sim \frac{P_{\rm syn}(t)}{4 \pi R^2_{\rm PWN}(t) c} \propto \eta (1 - \eta) t^{-5}.
\end{equation}
Then we obtain

\begin{equation}\label{eq27}
\frac{P_{\rm SSC}(t)}{P_{\rm syn}(t)} \sim \frac{f_{\rm KN} \xi(t)}{3} \cdot \frac{t}{\tau_{\rm syn}(\gamma_{\rm b}, t)} \cdot \frac{v_{\rm PWN}}{c} \cdot \frac{1 - \eta}{\eta} \propto (1- \eta ) t^{-2}.
\end{equation}
where $\tau_{\rm syn}(\gamma_{\rm b}, t) = N(\gamma_{\rm b}, t) \gamma^2_{\rm b} m_{\rm e} c^2 / P_{\rm syn}(t)$ is the synchrotron cooling time of the particles with $\gamma = \gamma_{\rm b}$ at time $t$.

All the powers $P_{\rm syn}(t) \propto \eta (1 - \eta) t^{-3}$, $P_{\rm IC/CMB}(t) \propto (1 - \eta)$, and $P_{\rm SSC}(t) \propto \eta (1 - \eta)^2 t^{-5}$ depend on the fraction parameter $\eta$ and time in different ways.
These general characteristics of the radiation powers make it possible to estimate the fraction parameter $\eta$ from the viewpoint of the spectral evolution.
For example, the time dependence of $P_{\rm SSC}(t)$ and $P_{\rm IC/CMB}(t)$ can explain that the SSC flux is dominant in high energy $\gamma$-rays in Figure \ref{f1}.
Using the same parameters as in Figure \ref{f1}, we find $P_{\rm IC/CMB}(t_{\rm{age}}) / P_{\rm syn}(t_{\rm{age}}) \sim 10^{-3}$ and $P_{\rm SSC}(t_{\rm{age}}) / P_{\rm syn}(t_{\rm{age}}) \sim 10^{-2}$.

In \citet{dd08}, they suggested that some of the TeV $\gamma$-ray sources without an X-ray counterpart can be old PWNe because the X-ray emitting particles are cooled more rapidly than the TeV $\gamma$-ray emitting particles by the synchrotron cooling.
In our model, old PWNe can also be the $\gamma$-ray sources with no or weak X-ray counterpart because the IC/CMB is almost time independent.
However the same result comes from the different reason. 
Because the number of the high energy particles is even increasing in our model as seen in Figure \ref{f3}, this result comes from the rapid decrease of the magnetic field.
PWNe which are old and has a small fraction parameter $\eta$ can be recognized as the $\gamma$-ray source without an X-ray counterpart.
Note that $P_{\rm IC/CMB}(t)$ slowly decreases with time, when we properly consider the cooling effects $\xi(t)$, as seen in Figure \ref{f2} at 3kyr and 10kyr.

\subsection{Parameters Characterizing Particle Injection}\label{other}
The fitted parameters $p_1$, $p_2$, $\gamma_{\rm max}$, $\gamma_{\rm b}$, and $\gamma_{\rm min}$ relate to the broken power-law injection in equation (\ref{eq4}).
The above parameters include the information about the acceleration at the termination shock and the physics of the pulsar wind and the pulsar magnetosphere.
We discuss about them in the framework of our model.

For the low and high energy power-law indices at injection $p_1 = 1.5$ and $p_2 = 2.5$, we determine these values to reproduce the observed radio and X-ray photon indices, $\Gamma_{\rm r}$ and $\Gamma_{\rm X}$.
Radio photon index $\Gamma_{r}$ is related to $p_1$, but X-ray photon index $\Gamma_{X}$ is not simply related to $p_2$ because of the synchrotron cooling.
Because it is difficult to make a large spectral break $\Gamma_{\rm X} - \Gamma_{\rm r} = \Delta \Gamma > 0.5$ from the single power-law injection \citep[e.g.][]{r09}, we adopt the broken power-law injection.

For the maximum energy $\gamma_{\rm max} = 7.0 \times 10^9$, we determine this value to reproduce the observed spectral break at $\sim 100 \rm{MeV}$.
The maximum energy $\gamma_{\rm max}$ is conventionally given by comparing the acceleration time with the cooling time or the size of the acceleration site with the Lamor radius.
Both conditions give $\gamma_{\rm max} \sim 10^{10} $ using the magnetic field strength $\sim 100 \mu \rm G$ and the size of the acceleration site $\sim 0.1 \rm pc$.
The fitted value is near the limit of theoretical expectation.

For the break energy $\gamma_{\rm b} = 6.0 \times 10^5$, we determine this value to reproduce the observed spectral break around optical wavelengths.
Although the KC model related $\gamma_{\rm b}$ to the bulk Lorentz factor of the pulsar wind immediately upstream the termination shock, our model does not allow this connection as discussed below.

Finally, for the minimum energy $\gamma_{\rm min} = 1.0 \times 10^2$, we determine this value to reproduce the flux of the observed minimum frequency at radio wavelengths.
Because $p_1 > 1$, the particles around $\gamma_{\rm min}$ determine the injection of the particle number as $\dot{N}_e(t) \sim Q_{\rm inj}(\gamma_{\rm min}, t) \gamma_{\rm min}$.
The particle number conservation outside the pulsar light cylinder leads to the particle number flux into the PWN is much larger than the Goldreich-Julian number flux $\dot{n}_{\rm GJ}$, which is the particle number flux from the pulsar polar cap.
In our model, the pair production multiplicity at an age of a thousand year $\kappa \sim (L(t) / \gamma_{\rm b} m_{\rm e} c^2) ( \gamma_{\rm b} / \gamma_{\rm min} )^{p_1 - 1} \sim 10^6$ from equations (\ref{eq4}) and (\ref{eq5}).
Theoretically, the multiplicity is estimated as $\kappa \sim 10^3 - 10^5$.
Our value $\kappa \sim 10^6$ is somewhat large.
The mean energy of the injected particles $L(t) / \dot{N}_e(t) \sim \gamma_{\rm b} m_{\rm e} c^2 ( \gamma_{\rm b} / \gamma_{\rm min} )^{- p_1 + 1}$ is significantly smaller than $\gamma_{\rm b} m_{\rm e} c^2$.

\citet{aa96} did not use the broken power-law injection.
They divided the particles inside the PWN into the low energy particles as the relic electrons and the high energy particles as the wind electrons.
This assumption can reduce the multiplicity, but the origin of the relic electrons becomes another problem.
It is difficult to discuss anything more about this problem from our spectral evolution model.
One thing what we should note is that the relation between the radio flux decrease and the magnetic field evolution is always kept, i.e., the relation is independent of how and when the radio emitting particles are injected.
This is because the low energy particles whose power-law index is $p_1$ do exist from the observation and their distribution hardly changes by the cooling effect, as discussed in section \ref{crab_evolution}.

\section{DISCUSSIONS AND CONCLUSIONS}\label{discussion_conclusion}
The evolutions of the magnetic field and the particle distribution determine the spectral evolution of a PWN.
The evolution of the particle distribution is affected by the assumptions of the magnetic field evolution model, uniformity of the PWN, particle injection spectrum, and expansion evolution of the PWN.
We discuss about the effects of these assumptions which are made in our model.
Finally, we summarize the conclusions of this paper.

\subsection{Discussion}\label{discussion}
Our model of the magnetic field evolution is somewhat ad hoc.
As discussed in section \ref{mag_evol}, however, the time dependence of the magnetic field strength $B \propto t^{-a}$ for $t > \tau_0$ seems to be in a range of $1.5 \le a \le 2.0$ from other theoretical considerations and we adopt $a = 1.5$.
Moreover, because our result of the radio flux decrease of the Crab Nebula is almost consistent with the observation, our model of the magnetic field evolution can be near the truth. 

For the assumption of the uniform PWN, many non-uniform structures have been observed, such as the filamentary structures and the spatial variations of photon indices.
However, for the calculation of the total spectrum of the PWN, the energetics of the PWN is important in the lowest order.
We consider that the assumption of the uniform PWN is reasonable for the calculation of the total spectral evolution of the PWN.

For the injection spectrum of the particle distribution, the acceleration of the particles is an unsolved problem and we adopt the broken power-law injection.
It should be noted that one of the important conclusions in our study that old PWNe can be observed as $\gamma$-ray sources with no or weak X-ray counterpart is not affected by the broken power-law assumption.
This is because low energy particles do not contribute to X-ray and high energy $\gamma$-ray emissions.

The use of the time independent parameters $\gamma_{\rm min}$, $\gamma_{\rm b}$ and $\gamma_{\rm max}$ can be improved as time dependent parameters, because the physical condition of pulsar wind termination shock may be change with the decrease of the spin-down power of the pulsar.
Considering the time dependence of $\gamma_{\rm max}(t)$, in \citet{vd06}, they use the condition $r_{\rm L}(t) < 0.5 r_{\rm s}$ with time dependent magnetic field, where $r_{\rm L}$ is the electrons' Larmor radius and $r_{\rm s}$ is the radius of the termination shock.
As discussed in section \ref{other}, our model satisfy this condition.
Both $\gamma_{\rm min}$ and $\gamma_{\rm b}$ are important parameters because these may include the information about the pulsar magnetosphere and the pulsar wind.
However, the time dependences of them are uncertain.
For simplicity, we used all of them as the time independent parameters in the present paper.

Constant velocity expansion is a good assumption for young PWNe, although the expansion of the PWN should be calculated by taking account for the environment of the PWN \citep[e.g.][]{get09}.
In our model, the magnetic field evolution explicitly depends on the expansion velocity $v_{\rm PWN}$ (see equation (\ref{eq8})).
To understand a little more about the effects of the expansion evolution, we study how the Crab Nebula would be observed in the context of the constant velocity expansion.
The Crab Nebula is one of the sources without observable SNR shell.
It may be because the surrounding interstellar medium is less dense than other PWNe with observable SNR shell.
If the Crab Nebula were in the different surroundings, the expansion velocity would also be affected.
That is, if it were in a less or more dense surroundings, the expansion velocity would be more rapid or slower.

\begin{figure}
\plotone{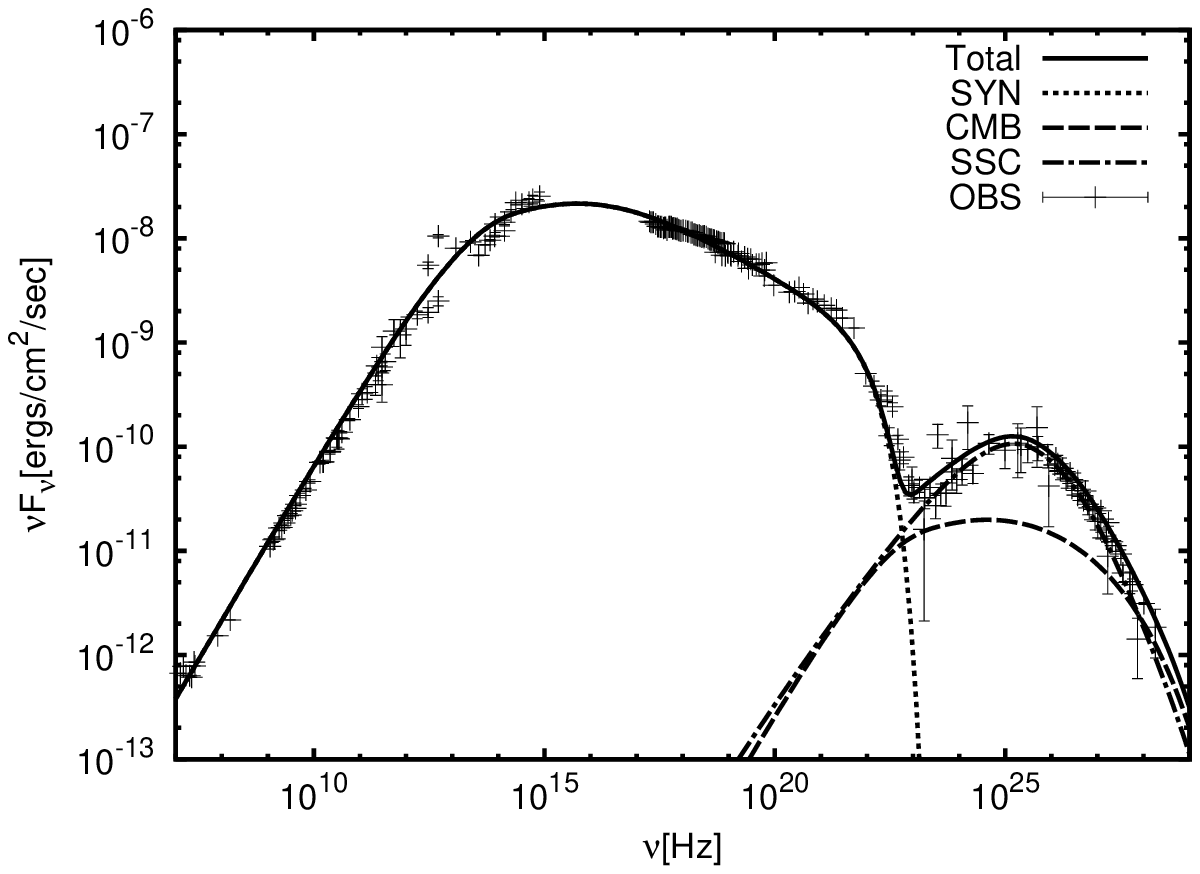}
\caption{Current spectrum of the Crab Nebula in our model and the observational data.
The solid line is the total spectrum which is the sum of the synchrotron (dotted line), IC/CMB (dashed line), and SSC (dot-dashed line) spectra, respectively.
The observed data taken from \citet[][]{bet77} (radio), \citet[][]{met10} (radio-optical), \citet[][]{g79, tet06, ns68} (IR), \citet[][]{ket01} (X-ray-$\gamma$-ray), \citet[][]{aet04, aet06, aet08, aet10} (very high energy $\gamma$-ray).
Used parameters are tabulated in Table\ref{tbl-1}. \label{f1}}
\end{figure}

\begin{figure}
\plotone{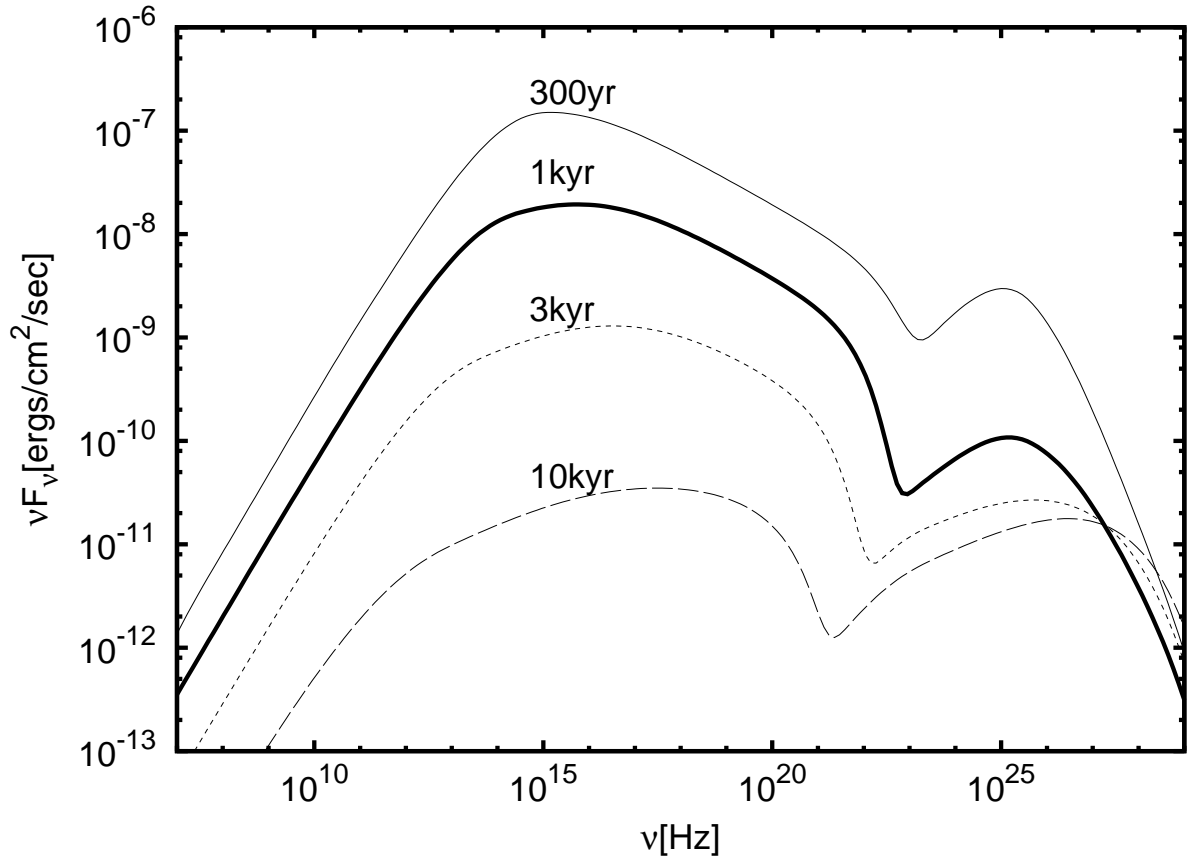}
\caption{Spectral evolution of the Crab Nebula.
The thin solid line is $300\rm yr$ from the birth.
The thick solid, thin dotted and thin dashed lines are $1\rm kyr$, $3\rm kyr$, and $10\rm kyr$ from birth, respectively.
Each line represents the total spectra which are the sum of the synchrotron, IC/CMB and SSC spectra.
Used parameters are the same as in Figure \ref{f1}.
\label{f2}}
\end{figure}

\begin{figure}
\plotone{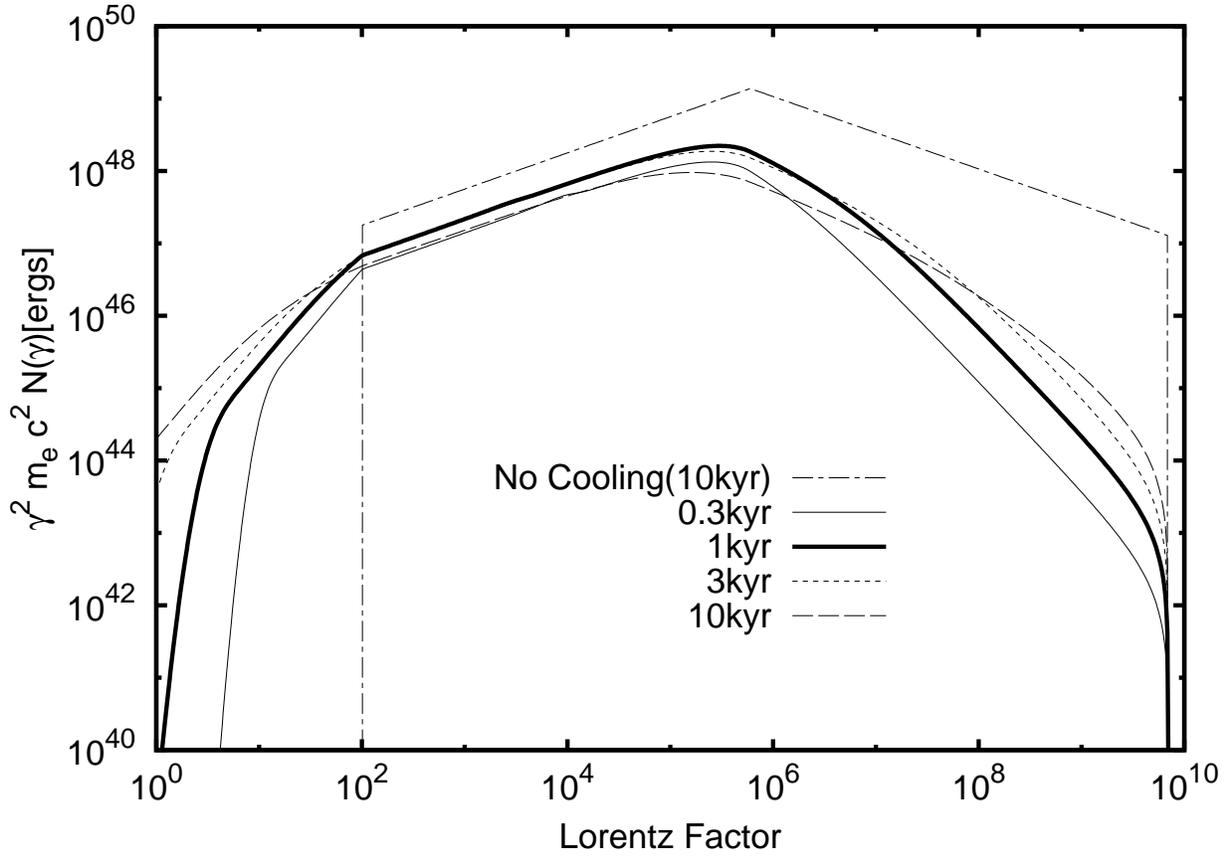}
\caption{Evolution of the particle distribution.
The thin solid line is the distribution at $300\rm yr$ from the birth.
The thick solid, thin dotted and thin dashed lines are those at $1\rm kyr$, $3\rm kyr$, and $10\rm kyr$, respectively.
The dot-dashed line is the total injected particles at an age of $10\rm kyr$.
Used parameters are the same as in Figure \ref{f1}.
\label{f3}}
\end{figure}

\begin{figure}
\plotone{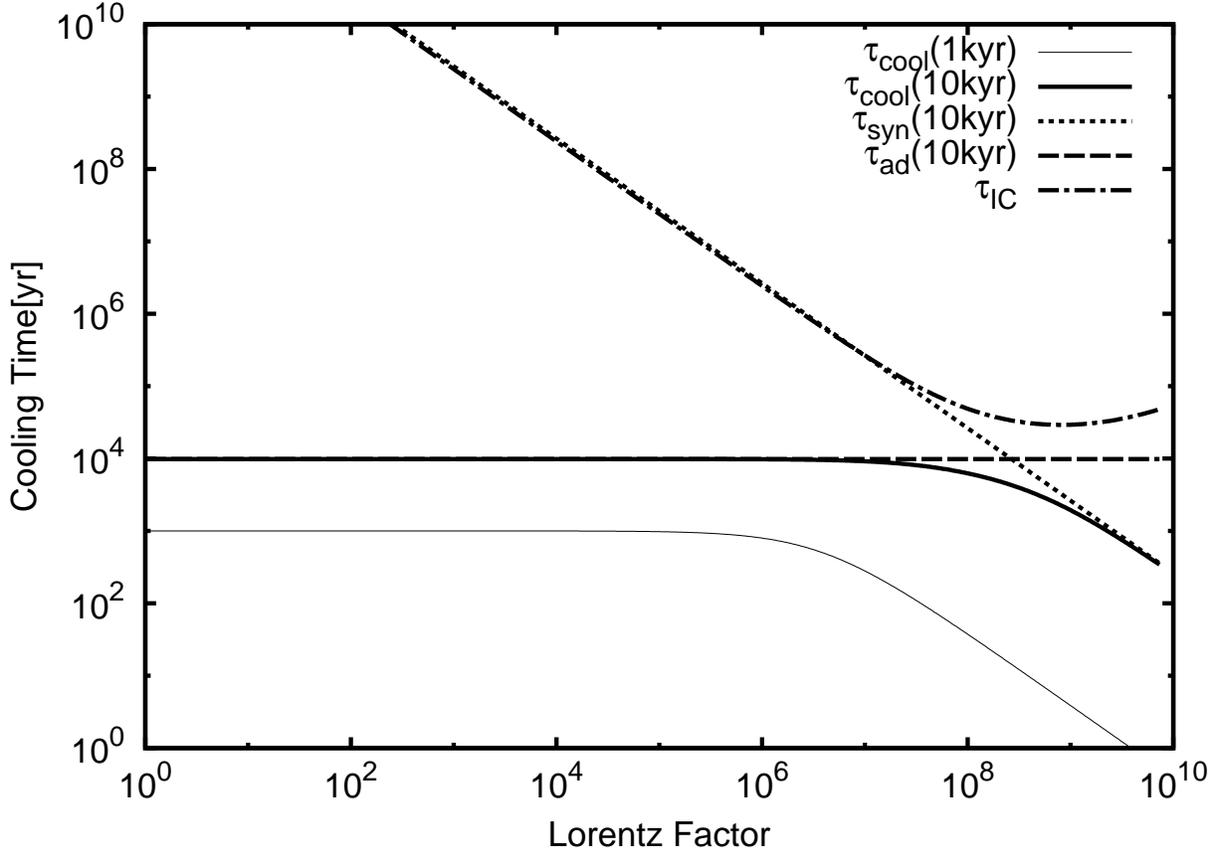}
\caption{Cooling times as a function of the Lorentz factor.
The thin and thick solid lines are the total cooling time $\tau_{\rm{cool}}(\gamma, t)$ at $t = 1\rm kyr$ and $10\rm kyr$, respectively.
The dotted, dashed and dot-dashed lines are $\tau_{\rm syn}(\gamma, 10\rm{kyr})$,  $\tau_{\rm ad}(\gamma, \rm{10kyr})$, and $\tau_{\mathrm{IC}}(\gamma)$, respectively.
Used parameters are the same as in Figure \ref{f1}.
\label{f4}}
\end{figure}

\begin{figure}
\plotone{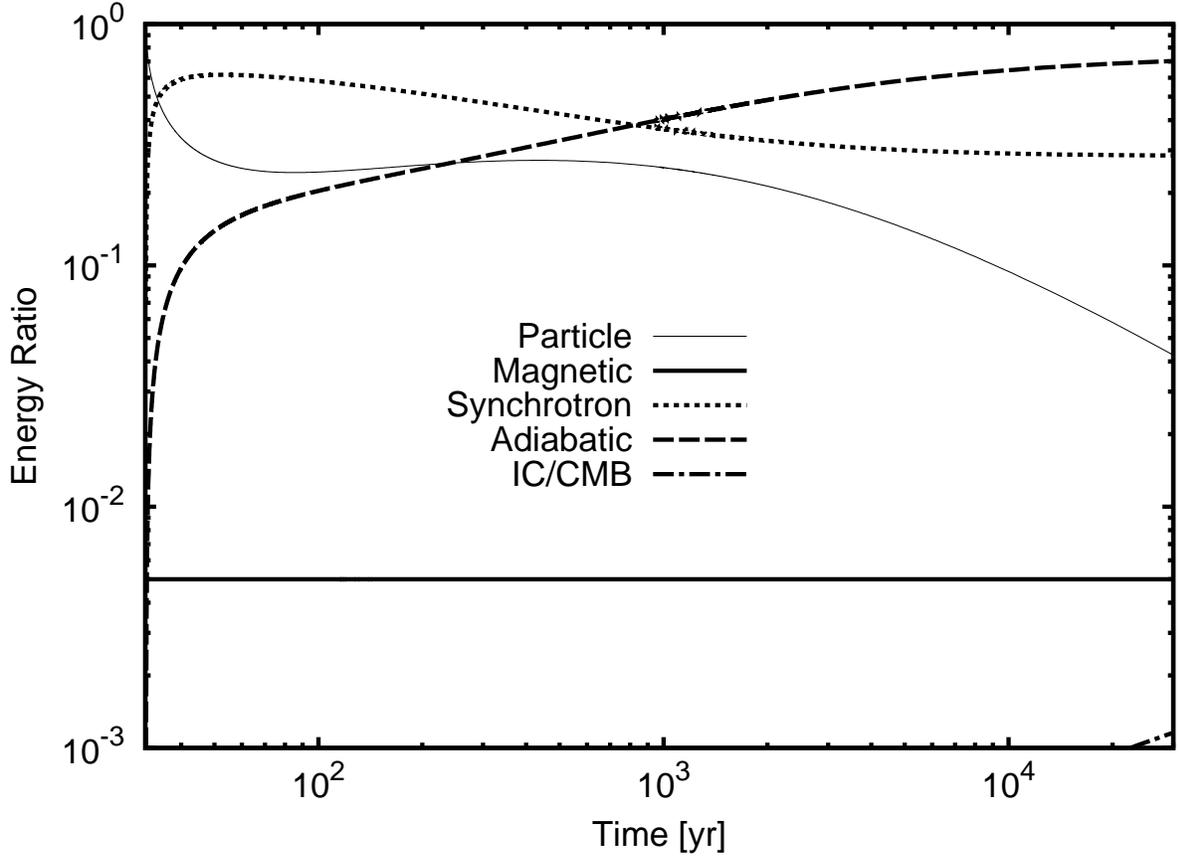}
\caption{Evolution of the energy content inside the Crab Nebula and the radiated energy and wasted energy by adiabatic expansion.
The thin solid line corresponds to the particle energy $\int^{\gamma_{\rm max}}_{\gamma_{\rm min}} \gamma m_{\rm e} c^2 N(\gamma, t) d\gamma $, the thick solid line is the magnetic field energy $(4 \pi / 3) R^3_{\rm PWN}(t) U_{\rm B}(t)$, the dotted, dashed and dot-dashed lines are the radiated energy via synchrotron radiation $\int^{\gamma_{\rm max}}_{\gamma_{\rm min}} \int^t_0 m_{\rm e} c^2 \vert \dot{\gamma}_{\rm syn}(\gamma, t') \vert d\gamma dt'$, the wasted energy via adiabatic expansion $\int^{\gamma_{\rm max}}_{\gamma_{\rm min}} \int^t_0 m_{\rm e} c^2 \vert \dot{\gamma}_{\rm ad}(\gamma, t') \vert d\gamma dt'$, and the radiated energy via inverse Compton scattering $\int^{\gamma_{\rm max}}_{\gamma_{\rm min}} \int^t_0 m_{\rm e} c^2 \vert \dot{\gamma}_{\rm IC}(\gamma) \vert d\gamma dt'$ respectively.
All the lines are normalized by the integrated spin-down energy $E_{\rm spin}(t)$.
Used parameters are the same as in Figure \ref{f1}.
\label{f5}}
\end{figure}

\begin{figure}
\plottwo{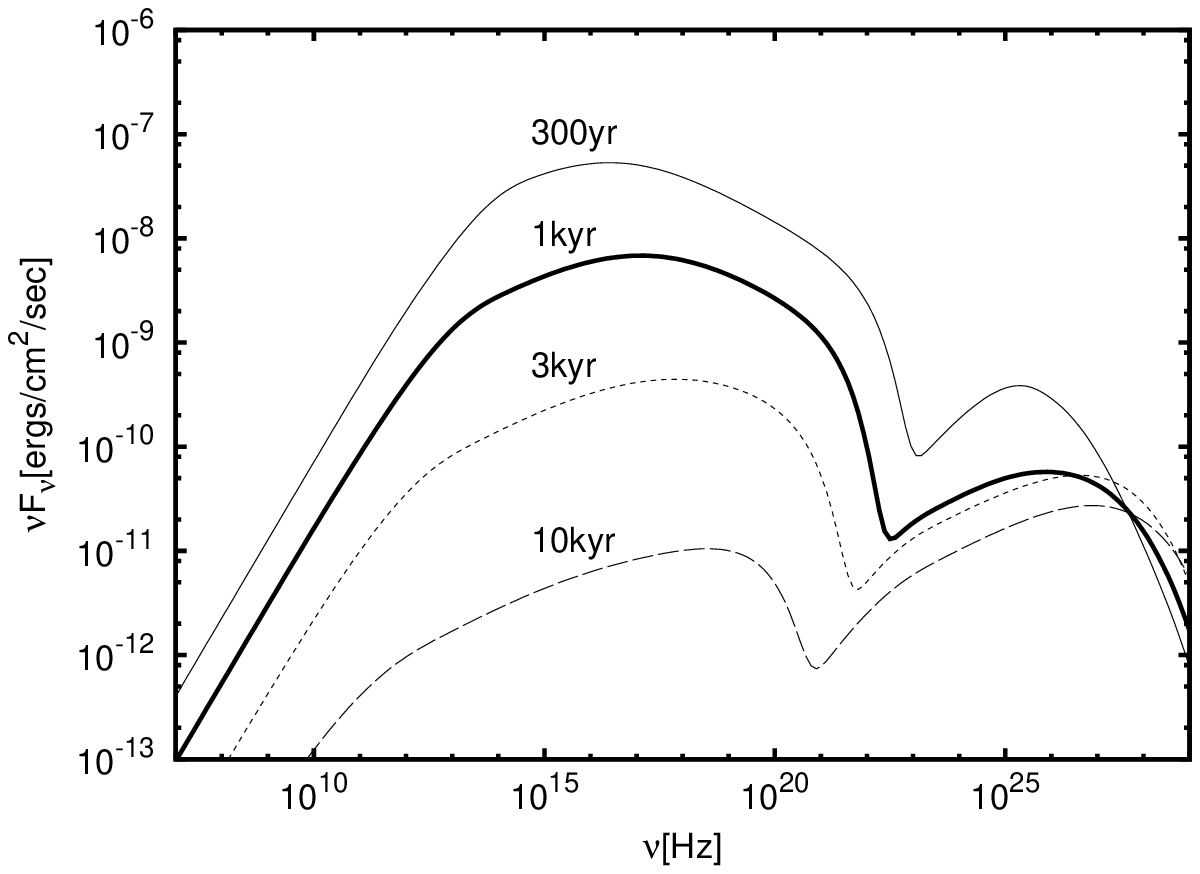}{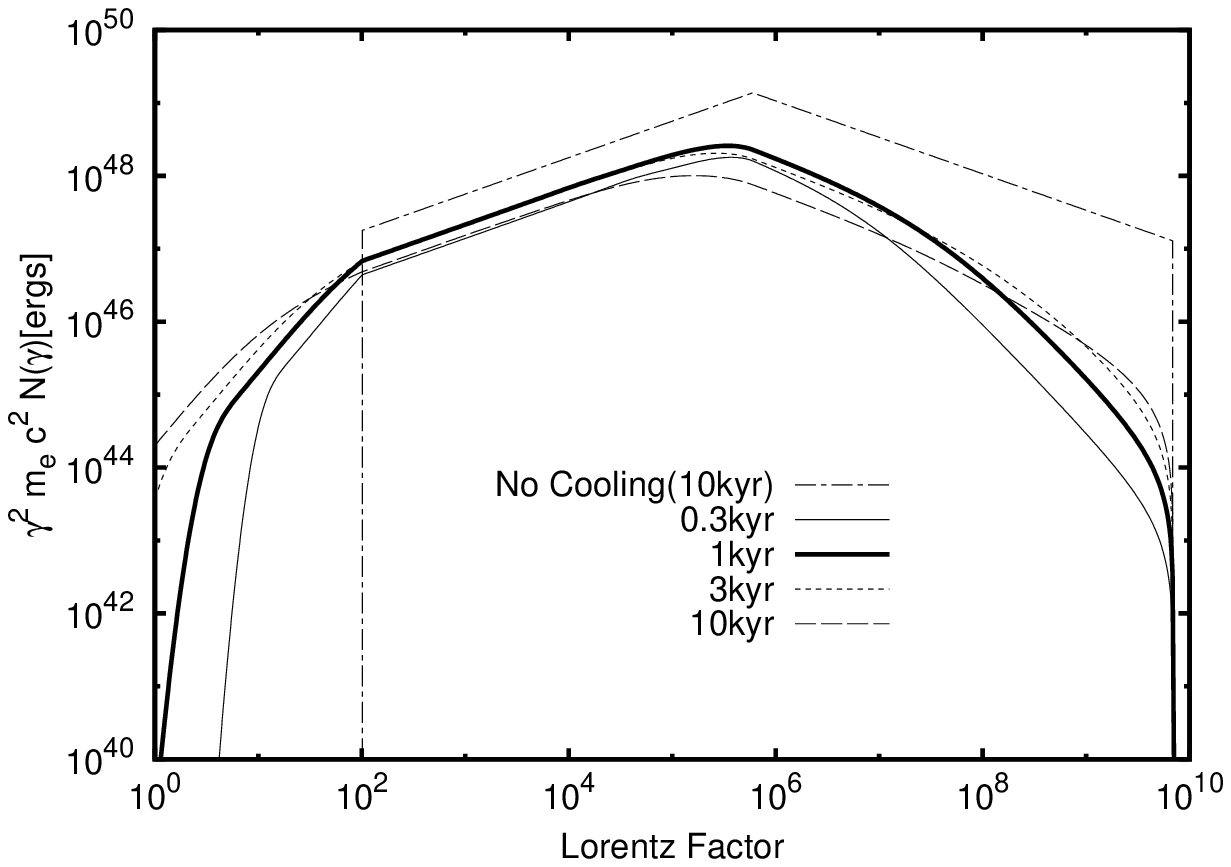}
\caption{Evolution of the emission spectrum (left panel) and particle distribution (right panel) of the PWN for the rapid expansion case.
Used parameters are the same as in Figure \ref{f1} except for the expansion velocity being twice ($v_{\rm PWN} = 3600 \rm km/s$).
\label{f6}}
\end{figure}

\begin{figure}
\plottwo{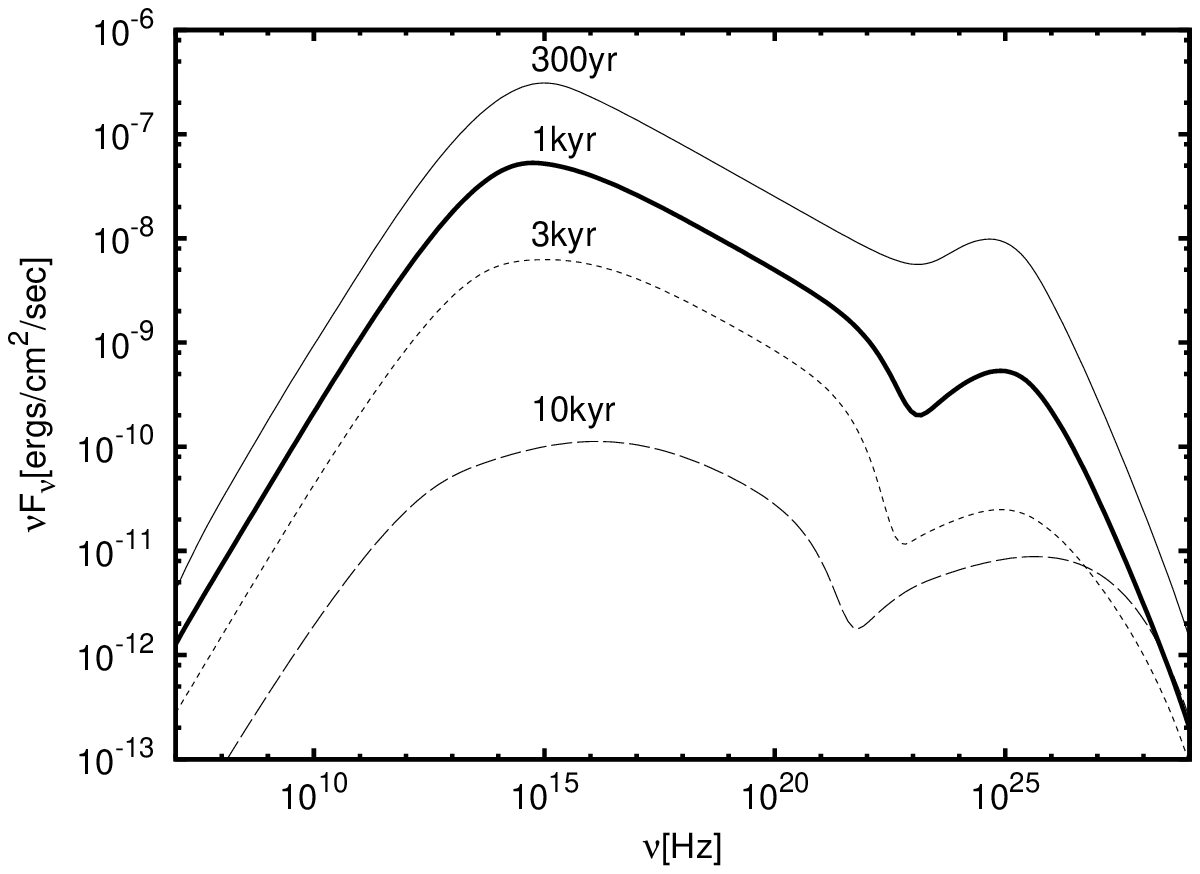}{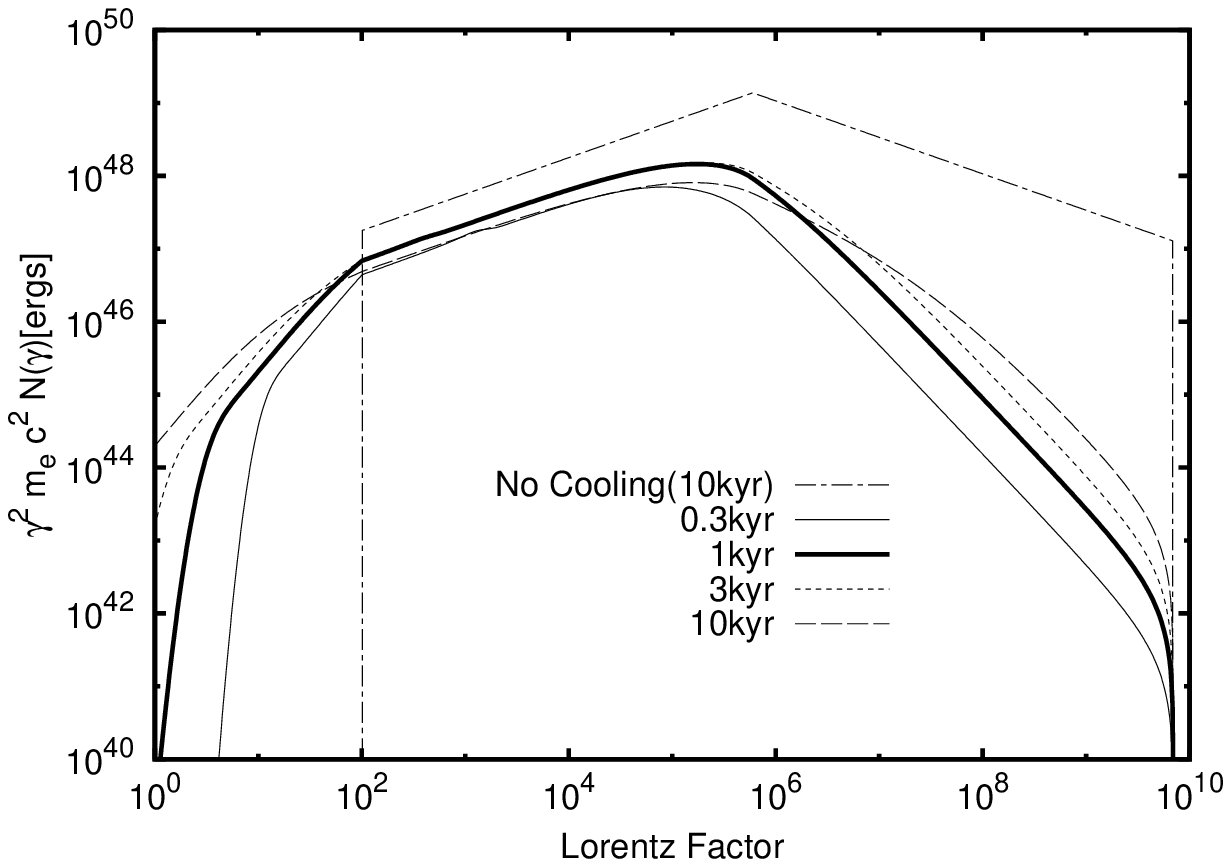}
\caption{Evolution of the emission spectrum (left panel) and particle distribution (right panel) of the PWN for the slow expansion case.
Used parameters are the same as in Figure \ref{f1} except for the expansion velocity being a half ($v_{\rm PWN} = 900 \rm km/s$).
\label{f7}}
\end{figure}

An example of a twice rapid expansion is shown in Figure \ref{f6}.
All the parameters except the expansion velocity are the same as in Figure \ref{f1}.
An example of a half velocity expansion is shown in Figure \ref{f7} and all parameters except for the expansion velocity are the same as in Figure \ref{f1}.
Note that both of them are hypothetical PWNe, not the Crab Nebula itself.
For the spectrum of the rapid expansion case, the absolute value of the flux becomes smaller and the flux ratio of the inverse Compton scattering to the synchrotron radiation is larger than the real Crab Nebula shown in Figure \ref{f1} and for the spectrum of the slow expansion case vice versa.
Comparing the particle distribution in Figure \ref{f6} with that in Figure \ref{f7}, the low energy particles take the same distribution, but the high energy particle distribution in Figure \ref{f7} is steeper than that in Figure \ref{f6}.
These spectral behaviors against the expansion velocity can be understood from equations (\ref{eq8}), (\ref{eq14}) and (\ref{eq16}).
In our model, the magnetic field becomes smaller when the expansion velocity becomes larger.
This leads to the difference in the absolute flux and the synchrotron cooling which changes the high energy particle distribution.
On the other hand, because the adiabatic cooling does not depend on the absolute value of the expansion velocity, the low energy particle distribution does not change.

Lastly, in \citet{aa96}, they included the infrared photons and the starlight for the target photon fields of the inverse Compton scattering.
Although these soft photons can significantly contribute to the $\gamma$-ray flux of other PWNe \citep[e.g.][]{pet06}, this is not the case of the Crab Nebula.
Because the Crab Nebula is located far away from the galactic center ($\sim 10 \rm kpc$) and galactic plane ($\sim 200 \rm pc$), the energy density of these soft photon fields is less than the solar neighborhood ($\sim 8 \rm kpc$ from the galactic center).
Even if we assume that the energy density of the infrared photons and the starlight is the same as the solar neighborhood, the inverse Compton scattering off these photon fields contributes less than 30 \% of the current total $\gamma$-ray flux.
Note that it also does not much affect the $\gamma$-ray spectrum, when the SSC flux decreases if the energy density of these soft photon fields is less than the half that in the solar neighborhood since inverse Compton scattering off the CMB dominates there.

\subsection{Conclusions}\label{conclusion}
In this paper, we built a model of the spectral evolution of PWNe and applied to the Crab Nebula as a calibrator of our model.
We solved the equation for the particle distribution function considering adiabatic and radiative losses with a simple model of magnetic field evolution.

The flux decrease of the $\gamma$-rays is more moderate than radio to X-rays, because the magnetic field decreases rapidly, which implies that old PWNe can be observed as $\gamma$-ray sources with no or weak X-ray counterpart.
Although \citet{dd08} obtained the same result but for a different reason that the X-ray emitting particles are cooled more rapidly than TeV $\gamma$-ray emitting particles.

The current observed spectrum of the Crab Nebula is reconstructed when the fraction parameter has a small value $\eta = 0.005$.
This is consistent with the prediction of the magnetization parameter $\sigma \ll 1$ obtained by \citet{kc84a}.
They obtained $\sigma \ll 1$ from the viewpoint of the current dynamical structure of the Crab Nebula, while we determine $\eta \ll 1$ from the viewpoint of the spectral evolution. 

The smaller value of the current magnetic field $B_{\rm now} = 85 \mu \rm G$ is needed to reconstruct the observed spectrum of the Crab Nebula.
This is consistent with that of \citet{vet08} for the spatially averaged magnetic field strength $\sim 100 \mu \rm G$ from their relativistic MHD simulation, but smaller than $\sim 300 \mu \rm G$ in most of other papers \citep[e.g.][]{aa96}.

Our model can predict the spectral evolution of the Crab Nebula, and the observed flux decrease of the Crab Nebula at radio wavelengths can be explained by our model.
This conclusion does not depend on the assumption of the broken power-law injection, and gives the validity of the our magnetic field evolution model.
The observed flux decrease of the Crab Nebula in optical wavelengths is somewhat larger than our model, but the trend that the decreasing rate increases with frequency matches observations.

The minimum energy $\gamma_{\rm min}$ is related to the pair production multiplicity in the pulsar magnetosphere, since low energy particles are assumed to be injected in the same way as high energy particles in our model.
Our result of the minimum energy $\gamma_{\rm min} = 1.0 \times 10^2$ and the low energy power-law index $p_1 = 1.5$ means that the multiplicity $\kappa \sim 10^6$ is necessarily larger than other models which adopt a separate origin of low energy particles. 

\acknowledgments
We are grateful to Y. Ohira for useful discussions.
This work is partly supported by KAKENHI (F. T. , 20540231)

\clearpage

\clearpage 

\clearpage 

\clearpage 

\clearpage 

\clearpage 

\clearpage 

\clearpage 

\begin{table}
\begin{center}
\caption{Used parameters to reproduce the current observed spectrum of the Crab Nebula.\label{tbl-1}}
\begin{tabular}{ccc}
\tableline\tableline
Adopted Parameter & Symbol & Value\\
\tableline
Current Period (s) & $\textit{P}$ & $3.31 \times 10^{-2}$\\
Current Period Derivative ($\rm s \cdot \rm s^{-1}$) & $\dot{\textit{P}}$ & $4.21 \times 10^{-13}$\\
Braking Index & $\textit{n}$ & 2.51\\
Age (yr) & $\textit{t}_{\rm age}$ & 950\\
Expansion Velocity (km/s) & $\textit{v}^{}_{\rm PWN}$ & 1800\\
\tableline
Fitted Parameter & & \\
\tableline
Fraction Parameter ($\dot{\textit{E}}_{\rm B} / (\dot{\textit{E}}_{\rm B} + \dot{\textit{E}}_{\rm e})  $)& $\eta$ & 0.005\\
Low Energy Power-law Index at Injection & $\textit{p}_1$ & 1.5\\
High Energy Power-law Index at Injection & $\textit{p}_2$ & 2.5\\
Maximum Energy at Injection & $\gamma_{\rm max}$ & $7.0 \times 10^9$\\
Break Energy at Injection & $\gamma_{\rm b}$ & $6.0 \times 10^5$\\
Minimum Energy at Injection & $\gamma_{\rm min}$ & $1.0 \times 10^2$\\
\tableline
\end{tabular}
\end{center}
\end{table}

\end{document}